\documentclass[aps,prd,nofootinbib,superscriptaddress,floatfix,nofootinbib,amsmath,amssymb]
{revtex4}
\usepackage[dvips]{graphicx}

\begin{document}

\title{Revisit of Tension in Recent SNIa Datasets}

\author{Miao Li}
\email{mli@itp.ac.cn} \affiliation{Institute of Theoretical Physics,
Chinese Academy of Sciences, Beijing 100080, China}
\affiliation{Kavli Institute for Theoretical Physics China,
Chinese Academy of Sciences, Beijing 100080, China}
\affiliation{Key Laboratory of Frontiers in Theoretical Physics,
Chinese Academy of Sciences, Beijing 100080, China}
\author{Xiao-Dong Li}
\email{renzhe@mail.ustc.edu.cn} \affiliation{Interdisciplinary Center for Theoretical Study,
University of Science and Technology of China, Hefei 230026, China}
\affiliation{Institute of Theoretical Physics,
Chinese Academy of Sciences, Beijing 100080, China}
\author{Shuang Wang}
\email{swang@mail.ustc.edu.cn} \affiliation{Department of Modern Physics,
University of Science and Technology of China, Hefei 230026, China}
\affiliation{Institute of Theoretical Physics,
Chinese Academy of Sciences, Beijing 100080, China}

\begin{abstract}
Although today there are many observational methods, Type Ia supernovae (SNIa) is still one of the most powerful tools to probe the mysterious dark energy (DE).
The most recent SNIa datasets are the 307 SNIa ``Union'' dataset \cite{kow08} and the 397 SNIa ``Constitution'' dataset \cite{hic09}.
In a recent work \cite{wei10}, Wei pointed out that both Union and Constitution datasets are in tension
with the observations of cosmic microwave background (CMB) and baryon acoustic oscillation (BAO),
and suggested that two truncated versions of Union and Constitution datasets, namely ``UnionT'' and ``ConstitutionT'', should be used to constrain various DE models.
But in \cite{wei10}, only the $\Lambda$CDM model is used to select the outliers from the Union and the Constitution dataset.
In principle, since different DE models may select different outliers, the truncation procedure should be performed for each different DE model.
In the present work, by performing the truncation procedure of \cite{wei10} for 10 different models,
we demonstrate that the impact of different models is negligible, and the approach adopted in \cite{wei10} is valid.
Moreover, by using the 4 SNIa datasets mentioned above, as well as the observations of CMB and BAO, we perform best-fit analysis on the 10 models.
It is found that: (1) For each DE model, the truncated SNIa datasets not only greatly reduce $\chi _{min}^{2}$ and $\chi _{min}^{2}/dof$,
but also remove the tension between SNIa data and other cosmological observations.
(2) The CMB data is very helpful to break the degeneracy among different parameters,
and plays a very important role in distinguishing different DE models.
(3) The current observational data are still too limited to distinguish all DE models.

\end{abstract}


\maketitle

\section{Introduction}

Observations of Type Ia supernovae (SNIa) \cite{rie98,per99,ton03,kno03,rie04a},
cosmic microwave background (CMB) \cite{ben03,spe03,spe07,pag07,hin07,kom09} and large scale structure (LSS) \cite{teg04a,teg04b,teg06}
all indicate the existence of dark energy (DE) driving the current accelerating expansion of the universe.
The most obvious theoretical candidate of DE is the cosmological constant $\Lambda$,
but it is plagued with the fine-tuning problem and the coincidence problem \cite{wein89,wein00,sah00,car01,pee03,pad03,cop06}.
There are also many dynamical DE models, such as quintessence \cite{pee88,rat88,rat88,zla99}, phantom \cite{cal02,car03}, $k$-essence \cite{arm99,chi00,arm01},
CPL \cite{che01,lin03}, tachyon \cite{pad02,bag03}, hessence \cite{wei05a,wei05b}, Chaplygin gas \cite{kam01}, generalized Chaplygin gas \cite{bto02},
holographic \cite{li04,hua05a,hua05b,li08,li09a,zhax10}, agegraphic \cite{wei08a,wei08b}, holographic Ricci \cite{gao09},
Yang-Mills condensate \cite{zhay07a,zhay07b,wans08a,wans08b}, etc.
Although numerous theoretical models have been proposed in the past decade, the nature of DE still remains a mystery.

In recent years, the numerical study of DE, i.e. utilizing cosmological observations to constrain DE models,
has become one of the most active fields in the modern cosmology
\cite{alb06,wany00,wany04,wany07,hut03,hut05,hua04,zhax04,zhax05,cha06,zhax07a,zhax07b,zhax09,wanb05,wanb06,ma09a,ma09b,wei07,wei08c,wei09}.
Although today there are many observational methods, SNIa is still one of the most powerful tools to probe the mysterious DE.
In the past decade, many SNIa datasets,
such as Gold04 \cite{rie04b}, Gold06 \cite{rie07}, SNLS \cite{ast07}, ESSENCE \cite{woo07}, Davis \cite{dav07}, have been released,
while the number and quality of SNIa have continually increased.
The most recent SNIa datasets are ``Union'' \cite{kow08} and ``Constitution'' \cite{hic09},
and they have been widely used in the literature
\cite{shaf09,li09b,qi09,hua09,wu09,chen09,gong10a,gong10b,san09,mor09,wans10}.
However, these SNIa datasets are not always consistent with other types of cosmological observations,
and are even in tension with other SNIa samples.
For examples, in \cite{jas05,jas06,ness05},
the Gold04 dataset was shown to be inconsistent with the SNLS dataset:
the SNLS dataset favors the $\Lambda$CDM model, while the Gold04 dataset favors the dynamical DE model.
In \cite{ness07}, by comparing the maximum likelihood fits of the CPL parameters ($w_0$, $w_1$) given by different SNIa samples,
Nesseris and Perivolaropoulos found that the Gold06 dataset is also in $2\sigma$ tension with the SNLS dataset.
Moreover, they also investigated how to remove this tension.
The method is simple.
First, they fitted the $\Lambda$CDM model to the whole 182 SNIa in the Gold06 dataset,
and obtained the best-fit parameter value of $\Lambda$CDM model (for SNIa data only).
Then, they calculated the relative deviation to the best-fit $\Lambda$CDM prediction, $|\mu_{obs}-\mu_{\Lambda CDM}|/\sigma_{obs}$, for all the 182 data points.
Here $\mu_{obs}$ is the observational value of distance modulus,
$\mu_{\Lambda CDM}$ is the theoretical value of distance modulus given by the best-fit $\Lambda$CDM model,
and $\sigma_{obs}$ is the 1$\sigma$ error of distance modulus.
By searching the SNIa samples satisfying $|\mu_{obs}-\mu_{\Lambda CDM}|/\sigma_{obs}>1.8$,
they isolated six SNIa that are mostly responsible for the tension.
Further, by using the random truncation method, they demonstrated that these 6 SNIa are systematically different from the Gold06 dataset.

In a recent work, by comparing the maximum likelihood fits of the
CPL parameters ($w_0$, $w_1$), Wei \cite{wei10} pointed out that
both Union and Constitution dataset are also in tension with the
observations of CMB and baryon acoustic oscillation (BAO). Moreover,
he also investigated how to remove these tensions. By using the
method of truncation of \cite{ness07}, Wei found out the main
sources that are responsible for the tensions: for the Union set,
there are 21 SNIa differing from the best-fit $\Lambda$CDM
prediction beyond $1.9\sigma$ (i.e. $|\mu_{obs}-\mu_{\Lambda
CDM}|/\sigma_{obs}>1.9$); and for the Constitution set, there are 34
SNIa differing from the best-fit $\Lambda$CDM prediction beyond
$1.9\sigma$. The specific limit of truncation (i.e. $1.9\sigma$) is
chosen based on two considerations: first, the tension between SNIa
samples and other observations can be completely removed; second,
the number of usable SNIa can be preserved as much as possible (see
\cite{wei10} for details). By subtracting these outliers from the
Union dataset and the Constitution dataset, respectively, two new
SNIa datasets, ``UnionT'' and ``ConstitutionT'' (``T'' stands for
``truncated''), were obtained. Further, Wei argued that the UnionT
and the ConstitutionT datasets are fully consistent with the other
cosmological observations, and should be used to constrain various
DE models. But in \cite{wei10}, only the $\Lambda$CDM model is used
to select the outliers from the Union and the Constitution dataset.
Since different DE models may select different outliers, one may
doubt whether the approach adopted in \cite{wei10} is valid. In
principle, the truncation procedure should be performed for each
different DE model. Only if the impact of different models is
negligible, one can conclude that the conclusion of Wei is correct.
So in the present work, we shall consider 10 different models. The
truncation procedure will be performed for all these 10 models, and
the corresponding cosmological consequences will be explored.

This paper is organized as follows:
In Section 2, we briefly describe 10 theoretical models considered in this work.
In Section 3, we present the method of data analysis, as well as the SNIa datasets we used in this paper.
By performing the truncation procedure of \cite{wei10} for 10 different models, we demonstrate that the approach adopted in \cite{wei10} is valid.
In Section 4, we show the data fitting results of 10 models, and present the corresponding conclusions.
Section 5 is a short summary.
In this work, we assume today's scale factor $a_{0}=1$, so the redshift $z$ satisfies $z=a^{-1}-1$;
the subscript ``0'' always indicates the present value of the corresponding quantity, and the unit with $c=\hbar=1$ is used.

\section{Models}

For a spatially flat (the assumption of flatness is motivated by the inflation scenario)
Friedmann-Robertson-Walker (FRW) universe with matter component $\rho_{m}$ and DE component $\rho_{de}$,
the Friedmann equation reads
\begin{equation} \label{Fried}
3M_{Pl}^2H^2=\rho_{m}+\rho_{de},
\end{equation}
where $M_{Pl}\equiv 1/\sqrt{8\pi G}$ is the reduced Planck mass,
and $H\equiv \dot{a}/a$ is the Hubble parameter.
Using this formula, one can easily get
\begin{equation} \label{Ez1}
E(z)\equiv H(z)/H_{0} =\left[\Omega_{m0}(1+z)^3+(1-\Omega_{m0})f(z)\right]^{1/2},
\end{equation}
where $H_{0}$ is the Hubble constant, $\Omega_{m0}$ is the present
fractional matter density, and key function $f(z)\equiv
\rho_{de}(z)/\rho_{de} (0)$ is given by the specific DE model.
Equivalently, we have
\begin{equation} \label{Ez2}
E(z)=\left(\Omega_{m0}(1+z)^3\over
1-\Omega_{de}\right)^{1/2},
\end{equation}
where $\Omega_{de}\equiv \frac{\rho_{de}}{\rho_{c}} = \frac{\rho_{de}}{3M_{Pl}^2H^2}$ is the fractional DE density.
Clearly, different DE model will give different $E(z)$.
In the following, we shall briefly describe the models considered in this work.

\subsection{Single-parameter models}

(1) The $\Lambda$CDM model: The DE density is always a constant,
i.e. the equation of state (EOS) $w=-1$, so
\begin{equation} \label{LCDM}
E(z)=\sqrt{\Omega_{m0}(1+z)^3+(1-\Omega_{m0})}.
\end{equation}

(2) The Dvali-Gabadadze-Porrati (DGP) model: For this modified
gravity model, the form of $E(z)$ is \cite{dav00}
\begin{equation} \label{DGP}
E(z)=\sqrt{\Omega_{m0}(1+z)^3+\Omega_{r_c}}+\sqrt{\Omega_{r_c}},
\end{equation}
where $\Omega_{r_c}$ is a constant, satisfies
\begin{equation}
\Omega_{r_c}=\left(\frac{1-\Omega_{m0}}{2}\right)^{2}.
\end{equation}

(3) The agegraphic dark energy (ADE): The DE density is
characterized by the conformal age $\eta$ of the universe
\cite{wei08b},
\begin{equation} \label{ade}
\rho_{de}=3n^{2}M_{Pl}^{2}\eta^{-2},~
and~~\eta \equiv \int \frac{dt}{a}=\int \frac{da}{a^{2}H},
\end{equation}
where $n$ is a positive constant.
The equation of motion for $\Omega_{de}$ is given by \cite{wei08b}
\begin{equation} \label{ADE}
\frac{d\Omega_{de}}{dz}=-\Omega_{de}\left(1-\Omega_{de}\right)
\left[3(1+z)^{-1}-\frac{2}{n}\sqrt{\Omega_{de}}\right].
\end{equation}
As in \cite{wei08b}, we choose the initial condition,
$\Omega_{de}(z_{ini})=n^2(1+z_{ini})^{-2}/4$, at $z_{ini}=2000$,
then Eq. (\ref{ADE}) can be numerically solved.
Substituting the results of Eq. (\ref{ADE}) into Eq. (\ref{Ez2}), the key function $E(z)$ can be obtained.
Notice that once $n$ is given, $\Omega_{m0}=1-\Omega_{de}(0)$ can be naturally obtained by solving Eq. (\ref{ADE}),
so the ADE model is a single-parameter model.

\subsection{Two-parameter models}

(4) The XCDM model: DE has a constant EOS $w$, then
\begin{equation} \label{XCDM}
E(z)=\sqrt{\Omega_{m0}(1+z)^3+(1-\Omega_{m0})(1+z)^{3(1+w)}}.
\end{equation}

(5) The Chaplygin gas (CG) model: The EOS of DE has the form
\cite{kam01}
\begin{equation} \label{cg}
p_{de}=-\frac{A}{\rho_{de}},
\end{equation}
where $A$ is a positive constant, and $p_{de}$ is the pressure of DE.
From this assumption, one can get \cite{kam01}
\begin{equation} \label{CG}
E(z)=\sqrt{\Omega_{m0}(1+z)^3+(1-\Omega_{m0})\sqrt{A_c+(1-A_c)(1+z)^6}},
\end{equation}
here $A_c=A/\rho_{de}(0)$ is also a positive constant.

(6) The holographic dark energy (HDE) model: The DE density is
characterized by the future event horizon $L$ of the universe
\cite{li04},
\begin{equation} \label{hde}
\rho_{de}=3c^{2}M_{Pl}^{2}L^{-2},~
and~~L = a\int_{t}^{\infty}\frac{dt^{\prime}}{a}=a\int_{a}^{\infty}\frac{da^{\prime}}{Ha^{\prime2}},
\end{equation}
where $c$ is a positive constant.
The equation of motion for $\Omega_{de}$ is given by \cite{li04}
\begin{equation} \label{HDE}
\frac{d \Omega_{de}}{dz}=-(1+z)^{-1}\Omega_{de}(1-\Omega_{de})\left(1+{2\over c}\sqrt{\Omega_{de}}\right).
\end{equation}
Solving  Eq. (\ref{HDE}) numerically and substituting the corresponding results into Eq. (\ref{Ez2}),
$E(z)$ can be obtained.

(7) The holographic Ricci dark energy (RDE) model: The DE density is
characterized by the Ricci scalar ${\cal R}$ \cite{gao09},
\begin{equation} \label{rde}
\rho_{de}=-{\alpha\over 16\pi G}{\cal R},~
and~~{\cal R}=-6\left(\dot{H}+2H^2\right),
\end{equation}
where $\alpha$ is a positive constant.
The form of $E(z)$ in this case is \cite{gao09}
\begin{equation} \label{RDE}
E(z)=\sqrt{\frac{2 \Omega_{m0}}{2-\alpha}(1+z)^{3}+(1-{2\Omega_{m0}\over 2 -\alpha})(1+z)^{(4-{2\over\alpha})}}.
\end{equation}

\subsection{Three-parameter models}

(8) The linear parameterization (LP) model: The EOS of DE is
parameterized as
\begin{equation} \label{lp1}
w=w_0+w_1z,
\end{equation}
where $w_0$ and $w_1$ are constants.
Then
\begin{equation} \label{lp2}
E(z)=\sqrt{\Omega_{m0}(1+z)^3+(1-\Omega_{m0})(1+z)^{3(1+w_0-w_1)}\exp(3w_1 z)}.
\end{equation}

(9) The Chevallier-Polarski-Linder (CPL) model: The EOS of DE is
parameterized as \cite{che01,lin03}
\begin{equation} \label{cpl1}
w=w_0+w_1\frac{z}{1+z},
\end{equation}
then
\begin{equation} \label{cpl2}
E(z)=\sqrt{\Omega_{m0}(1+z)^3+(1-\Omega_{m0})(1+z)^{3(1+w_0+w_1)}\exp\left(-\frac{3w_1 z}{1+z}\right)}.
\end{equation}

(10) The generalized Chaplygin gas (GCG) model: The EOS of DE has
the following form \cite{bto02}
\begin{equation} \label{gcg}
p_{de}=-\frac{A}{\rho_{de}^{\alpha}},
\end{equation}
where $\alpha$ is also a positive constant and $\alpha=1$ corresponds to the CG model.
One can get \cite{bto02}
\begin{equation} \label{GCG}
E(z)=\sqrt{\Omega_{m0}(1+z)^3+(1-\Omega_{m0})\left(A_s+(1-A_s)(1+z)^{3(1+\alpha)}\right)^{1/1+\alpha}}.
\end{equation}
here $A_s=A/\rho_{de}^{1+\alpha}(0)$.

\section{Methodology}

\subsection{Data analysis}
In this work we adopt $\chi^2$ statistics.
For a physical quantity $\xi$ with experimentally measured value $\xi_{obs}$,
standard deviation $\sigma_{\xi}$, and theoretically predicted value $\xi_{th}$,
the $\chi^2$ value is given by
\begin{equation}
\label{eq:chi2_xi}
\chi_{\xi}^2=\frac{\left(\xi_{th}-\xi_{obs}\right)^2}{\sigma_{\xi}^2}.
\end{equation}
The total $\chi^2$ is the sum of all $\chi_{\xi}^2$s,
i.e.
\begin{equation}
\label{eq:chi2}
\chi^2=\sum_{\xi}\chi_{\xi}^2.
\end{equation}

First, we consider the SNIa data that are given in terms of the distance modulus $\mu_{ obs}(z_i)$.
The theoretical distance modulus is defined as
\begin{equation}
\mu_{th}(z_i)\equiv 5 \log_{10} {D_L(z_i)} +\mu_0,
\end{equation}
where $\mu_0\equiv 42.38-5\log_{10}h$ with $h$ the Hubble constant $H_0$ in units of 100 km/s/Mpc,
and in a flat universe the Hubble-free luminosity distance $D_L\equiv H_0 d_L$ ($d_L$ denotes the physical luminosity distance) is
\begin{equation}
D_L(z)=(1+z)\int_0^z {dz'\over E(z'; \theta)},
\end{equation}
where $\theta$ denotes the model parameters.
The $\chi^2$ for the SNIa data is
\begin{equation}
\chi^2_{SN}(\theta)=\sum\limits_{i=1}{[\mu_{obs}(z_i)-\mu_{th}(z_i;\theta)]^2\over \sigma_i^2},
\label{ochisn}
\end{equation}
where $\mu_{obs}(z_i)$ and $\sigma_i$ are the observed value and the corresponding 1$\sigma$ error of distance modulus for each supernova, respectively.
Parameter $\mu_0$ is a nuisance parameter but it is independent of the data and the dataset.
Following \cite{peri05}, the minimization with respect to $\mu_0$
can be made trivially by expanding the $\chi^2$ of Eq. (\ref{ochisn}) with respect to $\mu_0$ as
\begin{equation}
\chi^2_{SN}(\theta)=A(\theta)-2\mu_0 B(\theta)+\mu_0^2 C,
\end{equation}
where
\begin{equation}
A(\theta)=\sum\limits_{i}{[\mu_{obs}(z_i)-\mu_{th}(z_i;\mu_0=0,\theta)]^2\over
\sigma_i^2},
\end{equation}
\begin{equation}
B(\theta)=\sum\limits_{i}{\mu_{obs}(z_i)-\mu_{th}(z_i;\mu_0=0,\theta)\over \sigma_i^2},
\end{equation}
\begin{equation}
C=\sum\limits_{i}{1\over \sigma_i^2}.
\end{equation}
Evidently, Eq. (\ref{ochisn}) has a minimum for $\mu_0=B/C$ at
\begin{equation} \label{tchi2sn}
\tilde{\chi}^2_{SN}(\theta)=A(\theta)-{B(\theta)^2\over C}.
\end{equation}
Since $\chi^2_{SN, min}=\tilde{\chi}^2_{SN,min}$,
instead of minimizing $\chi^2_{SN}$ we will minimize $\tilde{\chi}^2_{SN}$ which is independent of the nuisance parameter $\mu_0$.

Next, we consider constraints from the CMB and the LSS observations.
For the CMB data, we use the CMB shift parameter $R$, given by
\cite{bon97,wangy06}
\begin{equation}
R\equiv \Omega_{m0}^{1/2}\int_0^{z_{rec}}{dz'\over E(z')},
\end{equation}
where the redshift of recombination $z_{rec}=1091.3$ \cite{kom10}.
The measured value of $R$ has been updated to be $R_{obs}=1.725\pm 0.018$ from the WMAP7 observations \cite{kom10}.
For the LSS data, we use the BAO distance measurements obtained at $z=0.2$ and $z=0.35$ from the joint analysis of the 2dFGRS and SDSS data \cite{perc09}.
The BAO distance ratio $D_V(z=0.35)/D_V(z=0.20) = 1.736 \pm 0.065$ was shown in \cite{perc09} to be a relatively model independent quantity.
Here $D_V(z)$ is defined as
\begin{equation}
D_V(z_{BAO})= \bigg\lbrack~\frac{z_{BAO}}{H(z_{BAO})}~\bigg( \int_0^{z_{BAO}}
\frac{dz}{H(z)}~\bigg)^2~\bigg\rbrack^{1/3} .
\end{equation}

The total $\chi^2$ is given by
\begin{equation}
\chi^2=\tilde{\chi}_{SN}^2+\chi_{CMB}^2+\chi_{BAO}^2~,
\end{equation}
where $\tilde{\chi}_{ SN}^2$ is given by Eq. (\ref{tchi2sn}), and the latter two terms are defined as
\begin{equation}\label{chiCMB}
\chi^2_{CMB}=\frac{(R-1.725)^2}{0.018^2},
\end{equation}
and
\begin{equation}\label{chiLSS}
\chi^2_{BAO}=\frac{(D_V(0.35)/D_V(0.20)-1.736)^2}{0.065^2}.
\end{equation}
The model parameters yielding a minimal $\chi^{2}$ is favored by the
observations.

To compare different models, a statistical variable must be chosen.
The $\chi _{min}^{2}$ is the simplest one,
but it has difficulty to compare different models with different number of parameters.
In this work, we will use $\chi _{min}^{2}/dof$ as a model selection criterion,
where $dof$ is the degree of freedom defined as
\begin{equation}
\label{eq:dof}
dof\equiv N-k,
\end{equation}
here $N$ is the number of data, and $k$ is the number of free parameters.
This model selection criterion has been widely used in the literature.

\subsection{SNIa Datasets}

Although today there are many observational methods, SNIa is still one of the most powerful tools to probe the mysterious DE.
In the past decade, many SNIa datasets have been released, while the number and quality of SNIa have continually increased.
In 2008, the Union dataset \cite{kow08} was released.
It includes the large samples of SNIa from the HST, SNLS and ESSENCE,
and contains 307 samples: 250 high redshift SNIa ($z>0.2$) and 57 low redshift SNIa ($z\leq 0.2$).
In 2009, the Constitution set \cite{hic09} was released. It adds 90 low redshift samples.
These two SNIa datasets have been widely used in the literature.

In a recent work \cite{wei10}, by comparing the maximum likelihood
fits of the CPL parameters ($w_0$, $w_1$), Wei pointed out that both
Union and Constitution datasets are in tension not only with the
observations of CMB and BAO, but also with the other SNIa datasets.
Moreover, he also investigated how to remove these tensions. The
method of Wei is as follows. First, he fitted the $\Lambda$CDM model
to all the data points in the specific SNIa dataset, and obtained
the best-fit parameter value of $\Lambda$CDM model (for SNIa data
only). Then, he calculated the relative deviation to the best-fit
$\Lambda$CDM prediction, $|\mu_{obs}-\mu_{\Lambda
CDM}|/\sigma_{obs}$, for all the SNIa data points. By searching the
SNIa samples satisfying $|\mu_{obs}-\mu_{\Lambda
CDM}|/\sigma_{obs}>1.9$, Wei get the main sources that are
responsible for the tensions. For the Union set, there are 21 SNIa
differing from the best-fit $\Lambda$CDM prediction beyond
$1.9\sigma$, which is called ``UnionOut'' subset.
\begin{itemize}
\item {\bf UnionOut\ subset (21 SNIa):}\\
{\rm 1992bs,\ 1995ac,\ 1999bm,\ 1997o,\ 2001hu,\ 1998ba,\
04Pat,\ 05Red,\ 2002hr,\ 03D4au,}\\
{\rm 04D3cp,\ 03D1fc,\ 03D4dy,\ 03D1co,\ b010,\ d033,\
g050,\ g055,\ k430,\ m138,\ m226}
\end{itemize}
For the Constitution set, There are 34 SNIa differing from the
best-fit $\Lambda$CDM prediction beyond $1.9\sigma$, which is called
``ConstitutionOut'' subset.
\begin{itemize}
\item {\bf ConstitutionOut\ subset (34 SNIa):}\\
{\rm 1992bs,\ 1992bp,\ 1995ac,\ 1999bm,\ 1996t,\ 1997o,\
1995aq,\ 2001hu,\ 1998ba,\ 04Pat,\ 05Red,}\\
{\rm 2002hr,\ 03D4au,\ 04D3gt,\ 04D3cp,\ 03D4at,\ 03D1fc,\
04D3co,\ 03D4dy,\ 04D3oe,\ 04D1ak,}\\
{\rm 03D1co,\ b010,\ d033,\ f076,\ g050,\ k430,\ m138,\
m226,\ sn01cp,\ sn02hd,\ sn03ic,\ sn07ca,\ sn07R}
\end{itemize}
It should be mentioned that $1.9\sigma$ is chosen based on two considerations:
first, the tension between SNIa samples and other observations can be completely removed;
second, the number of usable SNIa can be preserved as much as possible.
After taking into account these two factors, Wei found that $1.9\sigma$ is most appropriate for the truncation procedure.
By subtracting these two subsets from the Union dataset and the Constitution dataset, respectively,
two new SNIa samples, ``UnionT'' and ``ConstitutionT'', were obtained.
It is clear that UnionT has 286 SNIa samples, and ConstitutionT has 363 SNIa samples.
Analyzing these two truncated datasets with CPL model,
Wei \cite{wei10} argued that they are fully consistent with the other cosmological observations.

But in \cite{wei10}, only the $\Lambda$CDM model is used to select the outliers from the Union and the Constitution dataset.
Since different DE models may select different outliers (i.e. different ``UnionOut'' and``ConstitutionOut'' Samples),
one may doubt whether the approach adopted in \cite{wei10} is valid.
In principle, the truncation procedure should be performed for each different DE model.
Only if the impact of different models is negligible, one can conclude that the conclusion of Wei is correct.
So in this work, we perform the truncation procedure for 10 different DE models.
As in \cite{wei10}, we choose $1.9\sigma$ as the selection criterion.
The results are shown in table \ref{UnionOut} and table \ref{ConstitutionOut}.
From these two tables, it is seen that the difference among the outlier samples given by various models are very small,
and the impact of different models is negligible.
Therefore, the approach adopted in \cite{wei10} is valid.
Since the difference among the outlier samples given by various models are very small,
for simplicity, in this work we just adopt the UnionT and the ConstitutionT SNIa samples given by \cite{wei10}.
To explore the corresponding cosmological consequences,
in the following we shall study 10 models by using 4 SNIa datasets: Union, UnionT, Constitution and ConstitutionT.

\begin{table} \caption{The ``UnionOut'' samples that satisfying $|\mu_{obs}-\mu_{model}|/\sigma_{obs}>1.9$, for each model.
Here ``+'' denotes adding a data point, while ``-'' denotes reducing a data point.}
\begin{center}
\label{UnionOut}
\begin{tabular}{cccc}
\hline\hline
~~~Model~~~ & ~~~Numbers of ``UnionOut'' samples for each model~~~ & ~~~Differences with the $\Lambda$CDM UnionOut subset~~~ \\
\hline\hline
~~~$\Lambda$CDM~~~ & ~~~$21$~~~ & ~~~None~~~ \\
\hline
~~~DGP~~~ & ~~~$21$~~~ & ~~~None~~~ \\
\hline
~~~ADE~~~ & ~~~$21$~~~ & ~~~None~~~ \\
\hline
~~~XCDM~~~ & ~~~$22$~~~ & ~~~+1999gd ~ +f076 ~ -05Red~~~ \\
\hline
~~~CG~~~ & ~~~$22$~~~ & ~~~+1999gd ~ +f076 ~ -05Red~~~ \\
\hline
~~~HDE~~~ & ~~~$22$~~~ & ~~~+1999gd ~ +f076 ~ -05Red~~~ \\
\hline
~~~RDE~~~ & ~~~$22$~~~ & ~~~+1999gd ~ +f076 ~ -05Red~~~ \\
\hline
~~~LP~~~ & ~~~$19$~~~ & ~~~+1999gd ~ +f076 ~ -1999bm ~ -05Red ~ -03D1co ~ -k430~~~ \\
\hline
~~~CPL~~~ & ~~~$20$~~~ & ~~~+1999gd ~ +f076 ~ -1999bm ~ -05Red ~ -03D1co~~~ \\
\hline
~~~GCG~~~ & ~~~$22$~~~ & ~~~+1999gd ~ +f076 ~ -05Red~~~ \\
\hline\hline
\end{tabular}
\end{center}
\end{table}

\begin{table} \caption{The ``ConstitutionOut'' samples that satisfying $|\mu_{obs}-\mu_{model}|/\sigma_{obs}>1.9$, for each model.
Here ``+'' denotes adding a data point, while ``-'' denotes reducing a data point.}
\begin{center}
\label{ConstitutionOut}
\begin{tabular}{cccc}
\hline\hline
~~~Model~~~ & ~~~Numbers of ``ConstitutionOut'' samples for each model~~~ & ~~~Difference with the $\Lambda$CDM ConstitutionOut subset~~~ \\
\hline\hline
~~~$\Lambda$CDM~~~ & ~~~$34$~~~ & ~~~None~~~ \\
\hline
~~~DGP~~~ & ~~~$34$~~~ & ~~~None~~~ \\
\hline
~~~ADE~~~ & ~~~$33$~~~ & ~~~-sn07R~~~ \\
\hline
~~~XCDM~~~ & ~~~$34$~~~ & ~~~None~~~ \\
\hline
~~~CG~~~ & ~~~$34$~~~ & ~~~+1997aj ~ -1992bp~~~ \\
\hline
~~~HDE~~~ & ~~~$34$~~~ & ~~~None~~~ \\
\hline
~~~RDE~~~ & ~~~$34$~~~ & ~~~None~~~ \\
\hline
~~~LP~~~ & ~~~$32$~~~ & ~~~+1997aj ~ -1992bp ~ -1999bm ~ -03D4at~~~ \\
\hline
~~~CPL~~~ & ~~~$33$~~~ & ~~~+1997aj ~ +sn02bf ~ -1992bp ~ -1999bm ~ -sn07R~~~ \\
\hline
~~~GCG~~~ & ~~~$34$~~~ & ~~~+1997aj ~ -1992bp~~~ \\
\hline\hline
\end{tabular}
\end{center}
\end{table}

\section{Results and Conclusions}

We will present the results of data fitting in this section.
Using the Union, the UnionT, the Constitution and the ConstitutionT dataset, respectively,
we list the $\chi_{min}^{2}$ and the $\chi_{min}^{2}/dof$ for those 10 models,
in table \ref{table1}, table \ref{table2}, table \ref{table3}, and table \ref{table4}.
Here ``SNIa'' means only SNIa data is used in the analysis,
``SNIa+BAO'' means both SNIa data and BAO data are used,
``SNIa+CMB'' means both SNIa data and CMN data are used,
and ``SNIa+BAO+CMB'' means all these three types of observational data are taken into account.
The main conclusions are summarized as follows.

\begin{table} \caption{The $\chi_{min}^{2}$ and the $\chi_{min}^{2}/dof$ (in the Parentheses) for the 10 models, where the Union dataset is used.}
\begin{center}
\label{table1}
\begin{tabular}{ccccc}
\hline\hline
~~~Model~~~ & ~~~SNIa~~~ & ~~~SNIa+BAO~~~  & ~~~SNIa+CMB~~~ &  ~~~SNIa+BAO+CMB~~~ \\
\hline\hline
~~~$\Lambda$CDM~~~ & ~~~$311.936~(1.019)$~~~ & ~~~$313.205~(1.020)$~~~ &  ~~~$312.424~(1.018)$~~~ &  ~~~$313.594~(1.018)$~~~ \\
\hline
~~~DGP~~~ & ~~~$313.026~(1.023)$~~~ & ~~~$314.319~(1.024)$~~~ &  ~~~$339.280~(1.105)$~~~ &  ~~~$341.584~(1.109)$~~~ \\
\hline
~~~ADE~~~ & ~~~$313.536~(1.025)$~~~ & ~~~$314.838~(1.026)$~~~ &  ~~~$327.216~(1.066)$~~~ &  ~~~$329.252~(1.069)$~~~ \\
\hline
~~~XCDM~~~ & ~~~$310.682~(1.019)$~~~ & ~~~$311.966~(1.020)$~~~ &  ~~~$312.224~(1.020)$~~~ &  ~~~$313.456~(1.021)$~~~ \\
\hline
~~~CG~~~ & ~~~$310.434~(1.018)$~~~ & ~~~$311.628~(1.018)$~~~ &  ~~~$311.921~(1.019)$~~~ &  ~~~$313.193~(1.020)$~~~ \\
\hline
~~~HDE~~~ & ~~~$310.827~(1.019)$~~~ & ~~~$312.149~(1.020)$~~~ &  ~~~$311.218~(1.017)$~~~ &  ~~~$312.481~(1.018)$~~~ \\
\hline
~~~RDE~~~ & ~~~$310.682~(1.019)$~~~ & ~~~$311.966~(1.020)$~~~ &  ~~~$312.323~(1.021)$~~~ &  ~~~$313.801~(1.022)$~~~ \\
\hline
~~~LP~~~ & ~~~$309.984~(1.020)$~~~ & ~~~$310.744~(1.019)$~~~ &  ~~~$310.691~(1.019)$~~~ &  ~~~$311.978~(1.020)$~~~ \\
\hline
~~~CPL~~~ & ~~~$310.091~(1.020)$~~~ & ~~~$310.896~(1.019)$~~~ &  ~~~$310.906~(1.019)$~~~ &  ~~~$312.258~(1.020)$~~~ \\
\hline
~~~GCG~~~ & ~~~$310.405~(1.021)$~~~ & ~~~$311.401~(1.021)$~~~ &  ~~~$311.925~(1.023)$~~~ &  ~~~$313.325~(1.024)$~~~ \\
\hline\hline
\end{tabular}
\end{center}
\end{table}

\begin{table} \caption{The $\chi_{min}^{2}$ and the $\chi_{min}^{2}/dof$ (in the Parentheses) for the 10 models, where the UnionT dataset is used.}
\begin{center}
\label{table2}
\begin{tabular}{ccccc}
\hline\hline
~~~Model~~~ & ~~~SNIa~~~ & ~~~SNIa+BAO~~~  & ~~~SNIa+CMB~~~ &  ~~~SNIa+BAO+CMB~~~ \\
\hline\hline
~~~$\Lambda$CDM~~~ & ~~~$204.568~(0.718)$~~~ & ~~~$205.945~(0.720)$~~~ &  ~~~$205.575~(0.719)$~~~ &  ~~~$206.794~(0.721)$~~~ \\
\hline
~~~DGP~~~ & ~~~$205.234~(0.720)$~~~ & ~~~$206.634~(0.722)$~~~ &  ~~~$226.987~(0.794)$~~~ &  ~~~$229.388~(0.799)$~~~ \\
\hline
~~~ADE~~~ & ~~~$205.560~(0.721)$~~~ & ~~~$206.963~(0.724)$~~~ &  ~~~$216.201~(0.756)$~~~ &  ~~~$218.305~(0.761)$~~~ \\
\hline
~~~XCDM~~~ & ~~~$204.058~(0.719)$~~~ & ~~~$205.449~(0.721)$~~~ &  ~~~$204.855~(0.719)$~~~ &  ~~~$206.207~(0.721)$~~~ \\
\hline
~~~CG~~~ & ~~~$204.050~(0.718)$~~~ & ~~~$205.365~(0.721)$~~~ &  ~~~$204.552~(0.718)$~~~ &  ~~~$205.930~(0.720)$~~~ \\
\hline
~~~HDE~~~ & ~~~$204.070~(0.719)$~~~ & ~~~$205.496~(0.721)$~~~ &  ~~~$204.228~(0.717)$~~~ &  ~~~$205.614~(0.719)$~~~ \\
\hline
~~~RDE~~~ & ~~~$204.058~(0.719)$~~~ & ~~~$205.449~(0.721)$~~~ &  ~~~$205.875~(0.722)$~~~ &  ~~~$207.498~(0.726)$~~~ \\
\hline
~~~LP~~~ & ~~~$204.055~(0.721)$~~~ & ~~~$205.347~(0.723)$~~~ &  ~~~$204.429~(0.720)$~~~ &  ~~~$205.739~(0.722)$~~~ \\
\hline
~~~CPL~~~ & ~~~$204.057~(0.721)$~~~ & ~~~$205.391~(0.723)$~~~ &  ~~~$204.063~(0.719)$~~~ &  ~~~$205.517~(0.721)$~~~ \\
\hline
~~~GCG~~~ & ~~~$204.053~(0.721)$~~~ & ~~~$205.366~(0.723)$~~~ &  ~~~$204.545~(0.720)$~~~ &  ~~~$206.090~(0.723)$~~~ \\
\hline\hline
\end{tabular}
\end{center}
\end{table}

\begin{table} \caption{The $\chi_{min}^{2}$ and the $\chi_{min}^{2}/dof$ (in the Parentheses) for the 10 models, where the Constitution dataset is used.}
\begin{center}
\label{table3}
\begin{tabular}{ccccc}
\hline\hline
~~~Model~~~ & ~~~SNIa~~~ & ~~~SNIa+BAO~~~  & ~~~SNIa+CMB~~~ &  ~~~SNIa+BAO+CMB~~~ \\
\hline\hline
~~~$\Lambda$CDM~~~ & ~~~$465.604~(1.176)$~~~ & ~~~$466.902~(1.176)$~~~ &  ~~~$466.316~(1.175)$~~~ &  ~~~$467.525~(1.175)$~~~ \\
\hline
~~~DGP~~~ & ~~~$466.122~(1.177)$~~~ & ~~~$467.433~(1.177)$~~~ &  ~~~$498.264~(1.255)$~~~ &  ~~~$500.368~(1.257)$~~~ \\
\hline
~~~ADE~~~ & ~~~$466.275~(1.177)$~~~ & ~~~$467.580~(1.178)$~~~ &  ~~~$483.675~(1.218)$~~~ &  ~~~$485.585~(1.220)$~~~ \\
\hline
~~~XCDM~~~ & ~~~$465.602~(1.179)$~~~ & ~~~$466.901~(1.179)$~~~ &  ~~~$465.657~(1.176)$~~~ &  ~~~$466.947~(1.176)$~~~ \\
\hline
~~~CG~~~ & ~~~$465.293~(1.178)$~~~ & ~~~$466.564~(1.178)$~~~ &  ~~~$465.591~(1.176)$~~~ &  ~~~$466.891~(1.176)$~~~ \\
\hline
~~~HDE~~~ & ~~~$465.769~(1.179)$~~~ & ~~~$467.080~(1.179)$~~~ &  ~~~$466.030~(1.177)$~~~ &  ~~~$467.384~(1.177)$~~~ \\
\hline
~~~RDE~~~ & ~~~$465.602~(1.179)$~~~ & ~~~$466.901~(1.179)$~~~ &  ~~~$472.841~(1.194)$~~~ &  ~~~$474.466~(1.195)$~~~ \\
\hline
~~~LP~~~ & ~~~$461.071~(1.170)$~~~ & ~~~$461.570~(1.169)$~~~ &  ~~~$465.610~(1.179)$~~~ &  ~~~$466.910~(1.179)$~~~ \\
\hline
~~~CPL~~~ & ~~~$461.526~(1.171)$~~~ & ~~~$462.112~(1.170)$~~~ &  ~~~$465.636~(1.179)$~~~ &  ~~~$466.902~(1.179)$~~~ \\
\hline
~~~GCG~~~ & ~~~$465.080~(1.180)$~~~ & ~~~$466.360~(1.181)$~~~ &  ~~~$465.920~(1.180)$~~~ &  ~~~$466.895~(1.179)$~~~ \\
\hline\hline
\end{tabular}
\end{center}
\end{table}

\begin{table} \caption{The $\chi_{min}^{2}$ and the $\chi_{min}^{2}/dof$ (in the Parentheses) for the 10 models, where the ConstitutionT dataset is used.}
\begin{center}
\label{table4}
\begin{tabular}{ccccc}
\hline\hline
~~~Model~~~ & ~~~SNIa~~~ & ~~~SNIa+BAO~~~  & ~~~SNIa+CMB~~~ &  ~~~SNIa+BAO+CMB~~~ \\
\hline\hline
~~~$\Lambda$CDM~~~ & ~~~$269.081~(0.743)$~~~ & ~~~$270.610~(0.745)$~~~ &  ~~~$271.423~(0.748)$~~~ &  ~~~$272.764~(0.749)$~~~ \\
\hline
~~~DGP~~~ & ~~~$269.443~(0.744)$~~~ & ~~~$270.979~(0.747)$~~~ &  ~~~$292.067~(0.805)$~~~ &  ~~~$294.360~(0.809)$~~~ \\
\hline
~~~ADE~~~ & ~~~$269.613~(0.745)$~~~ & ~~~$271.139~(0.747)$~~~ &  ~~~$280.090~(0.772)$~~~ &  ~~~$282.147~(0.775)$~~~ \\
\hline
~~~XCDM~~~ & ~~~$269.066~(0.745)$~~~ & ~~~$270.599~(0.748)$~~~ &  ~~~$269.245~(0.744)$~~~ &  ~~~$270.754~(0.746)$~~~ \\
\hline
~~~CG~~~ & ~~~$268.992~(0.745)$~~~ & ~~~$270.502~(0.747)$~~~ &  ~~~$269.066~(0.743)$~~~ &  ~~~$270.594~(0.745)$~~~ \\
\hline
~~~HDE~~~ & ~~~$269.113~(0.745)$~~~ & ~~~$270.664~(0.748)$~~~ &  ~~~$269.130~(0.743)$~~~ &  ~~~$270.698~(0.746)$~~~ \\
\hline
~~~RDE~~~ & ~~~$269.066~(0.745)$~~~ & ~~~$270.599~(0.748)$~~~ &  ~~~$273.305~(0.755)$~~~ &  ~~~$275.180~(0.758)$~~~ \\
\hline
~~~LP~~~ & ~~~$268.812~(0.747)$~~~ & ~~~$270.021~(0.748)$~~~ &  ~~~$269.068~(0.745)$~~~ &  ~~~$270.603~(0.748)$~~~ \\
\hline
~~~CPL~~~ & ~~~$268.900~(0.747)$~~~ & ~~~$270.147~(0.748)$~~~ &  ~~~$269.122~(0.745)$~~~ &  ~~~$270.676~(0.748)$~~~ \\
\hline
~~~GCG~~~ & ~~~$268.980~(0.747)$~~~ & ~~~$270.473~(0.749)$~~~ &  ~~~$269.069~(0.745)$~~~ &  ~~~$270.599~(0.748)$~~~ \\
\hline\hline
\end{tabular}
\end{center}
\end{table}

(1) For each DE model, the truncated SNIa datasets not only greatly reduce $\chi _{min}^{2}$ and $\chi _{min}^{2}/dof$,
but also remove the tension between SNIa data and other cosmological observations.

Since it is too prolix to describe the results of all 10 models, in
the following we will just discuss the HDE model as an example. From
table \ref{table1} and table \ref{table2}, we can see that the
combined Union+BAO+CMB data gives a $\chi_{min}^{2}=312.481$ and a
$\chi_{min}^{2}/dof=1.018$, while the combined UnionT+BAO+CMB data
gives a $\chi_{min}^{2}=205.614$ and a $\chi_{min}^{2}/dof=0.719$.
This means that by dropping the 21 outliers, The $\chi_{min}^{2}$
and the $\chi_{min}^{2}/dof$ of the HDE model can reduce 106.867 and
0.299, respectively. From table \ref{table3} and table \ref{table4},
similar results are obtained. The combined Constitution+BAO+CMB data
gives a $\chi_{min}^{2}=467.384$ and a $\chi_{min}^{2}/dof=1.177$,
while the combined ConstitutionT+BAO+CMB data gives a
$\chi_{min}^{2}=270.698$ and a $\chi_{min}^{2}/dof=0.746$. This
means that by dropping the 34 outliers, The $\chi_{min}^{2}$ and the
$\chi_{min}^{2}/dof$ of the HDE model can reduce 196.686 and 0.431,
respectively. It should be pointed out that for all these 10 models,
the UnionT and the ConstitutionT dataset can greatly reduce the
corresponding $\chi _{min}^{2}$ and $\chi _{min}^{2}/dof$. Moreover,
the decreasing margin of $\chi _{min}^{2}$ and $\chi _{min}^{2}/dof$
for these 10 DE models are almost same.

To further verify this conclusion, we also adopt the method of random truncation used in \cite{ness07,wei10}.
The method is very simple.
First, we random select the outliers from the full Union dataset and the full Constitution dataset, respectively.
Notice that the numbers of data points in these random ``UnionOut'' subsets are same as that in the original UnionOut samples,
while the numbers of data points in these random ``ConstitutionOut'' subsets are same as that in the original ConstitutionOut samples.
As in \cite{ness07,wei10}, 500 random UnionOut subsets and 500 random ConstitutionOut subsets are selected.
By subtracting these outliers from the Union dataset and the Constitution dataset, respectively,
500 random UnionT subsets and 500 random ConstitutionT subsets are obtained.
Then, we perform best-fit analysis on the HDE model by using these random truncated SNIa subsets.
Using the SNIa data only,
we get the Mean of $\chi_{min}^{2}$ for 500 random UnionOut subsets and 500 random ConstitutionOut subsets, respectively.
Notice that the full Union set gives a $\chi_{min}^{2}=310.827$ and the full Constitution set gives a $\chi_{min}^{2}=465.769$,
the differences between the $\chi_{min}^{2}$ of the full SNIa dataset and the Mean of $\chi_{min}^{2}$ of these random truncated SNIa subsets are also obtained.
Next, we make a comparison for the original truncated SNIa subset and the random truncated SNIa subsets.
The results are shown in table \ref{table5}.
It is found that the $\chi_{min}^{2}$ of the HDE model can reduce 5 by dropping 1 data point in the original truncated SNIa subset,
but can only reduce 1 by dropping 1 data point in the random truncated SNIa subset.
Therefore, the original UnionOut and ConstitutionOut subsets are systematically different from the full Union and Constitution datasets,
and the original truncated supernova datasets provide a significantly better model fitting than the full SNIa datasets.

\begin{table} \caption{A comparison for the original truncated SNIa subset and the random truncated SNIa subsets.
Notice that the full Union set gives a $\chi_{min}^{2}=310.827$ and the full Constitution set gives a $\chi_{min}^{2}=465.769$.
Here $\Delta N$ is the number of data points in outliers,
$\overline{{\chi^2}}$ is the Mean of $\chi_{min}^{2}$ for those random truncated SNIa subsets,
and $\Delta \chi^2$ is the differences between the $\chi_{min}^{2}$ of the full SNIa dataset and the $\chi_{min}^{2}$ (or the Mean) of the random truncated SNIa subsets.}
\begin{center}
\label{table5}
\begin{tabular}{ccccc}
\hline\hline
~~~The subset of outliers~~~ & ~~~$\Delta N$~~~ & ~~~$\chi_{min}^{2}$ or $\overline{{\chi^2}}$~~~ & ~~~$\Delta \chi^2$~~~ & ~~~$\Delta \chi^2/\Delta N$~~~ \\
\hline\hline
~~~The original UnionOut subset~~~ & ~~~21~~~ & ~~~$204.070$~~~ & ~~~$106.757$~~~ & ~~~$5.084$~~~ \\
\hline
~~~500 random UnionOut subsets~~~ & ~~~21 (Mean)~~~ & ~~~$289.617$ (Mean)~~~ & ~~~$21.210$ (Mean)~~~ & ~~~$1.010$ (Mean)~~~ \\
\hline
~~~The original ConstitutionOut subset~~~ & ~~~34~~~ & ~~~$269.113$~~~ & ~~~$199.656$~~~ & ~~~$5.784$~~~ \\
\hline
~~~500 random ConstitutionOut subsets~~~ & ~~~34 (Mean)~~~ & ~~~$425.763$ (Mean)~~~ & ~~~$40.006$ (Mean)~~~ & ~~~$1.177$ (Mean)~~~ \\
\hline\hline
\end{tabular}
\end{center}
\end{table}

\begin{figure}
\includegraphics[scale=0.7, angle=0]{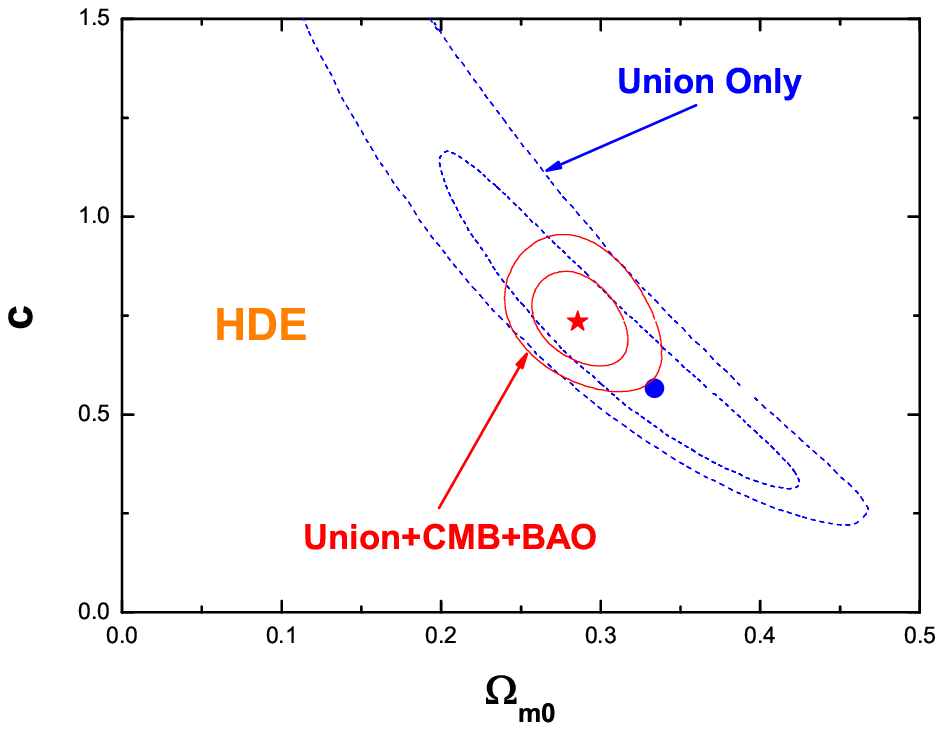}
\includegraphics[scale=0.7, angle=0]{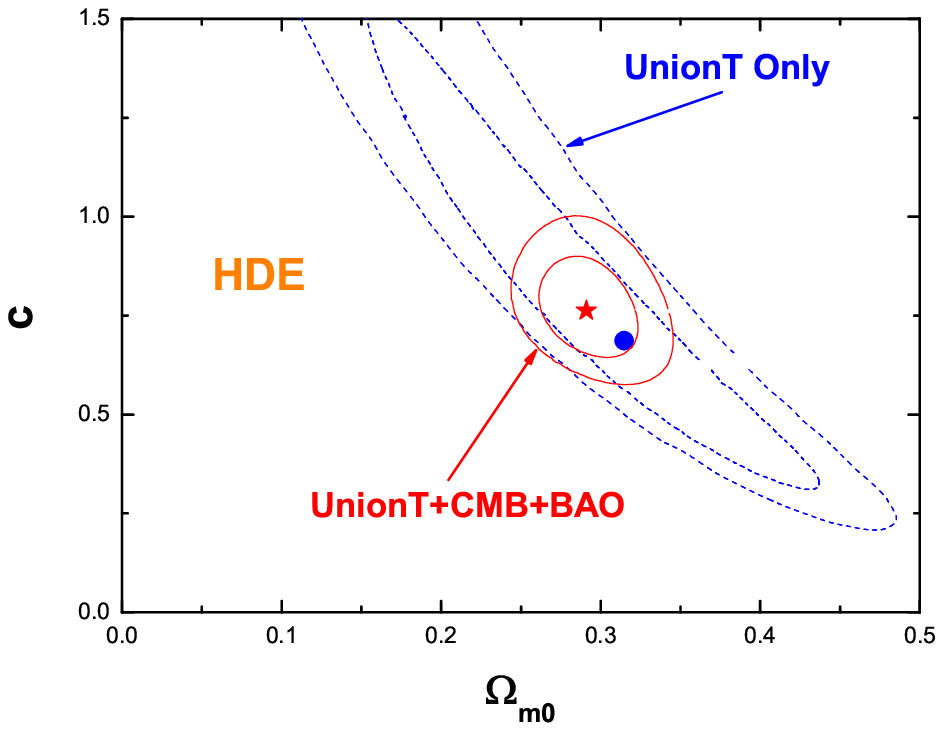}
\caption{\label{fig1} The $1\sigma$ and the $2\sigma$ CL contours for the HDE model.
The left panel is plotted by using the Union dataset, while the right panel is plotted by using the UnionT dataset.
For both these two panels, the blue dashed lines correspond to the constraints given by the SNIa data only,
and the red solid lines correspond to the constraints given by the combined SNIa+BAO+CMB data.
Moreover, we also plot the best-fit point for the SNIa data (blue point) and the best-fit point for the combined SNIa+BAO+CMB data (red star).
It should be pointed out that,
in the left panel, the best-fit point for the SNIa data is outside the $2\sigma$ confidence region given by the combined SNIa+BAO+CMB data,
while in the right panel, the best-fit point for the SNIa data is inside the $1\sigma$ confidence region given by the combined SNIa+BAO+CMB data.
This means that the UnionT dataset is very useful to remove the tension between SNIa data and other cosmological observations.}
\end{figure}

\begin{figure}
\includegraphics[scale=0.7, angle=0]{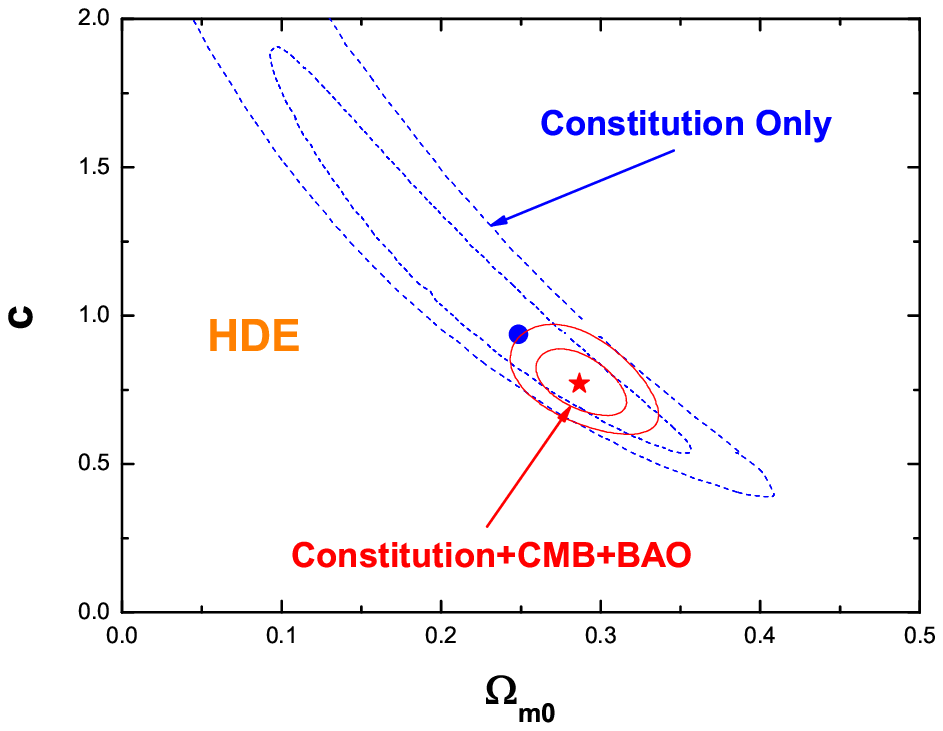}
\includegraphics[scale=0.7, angle=0]{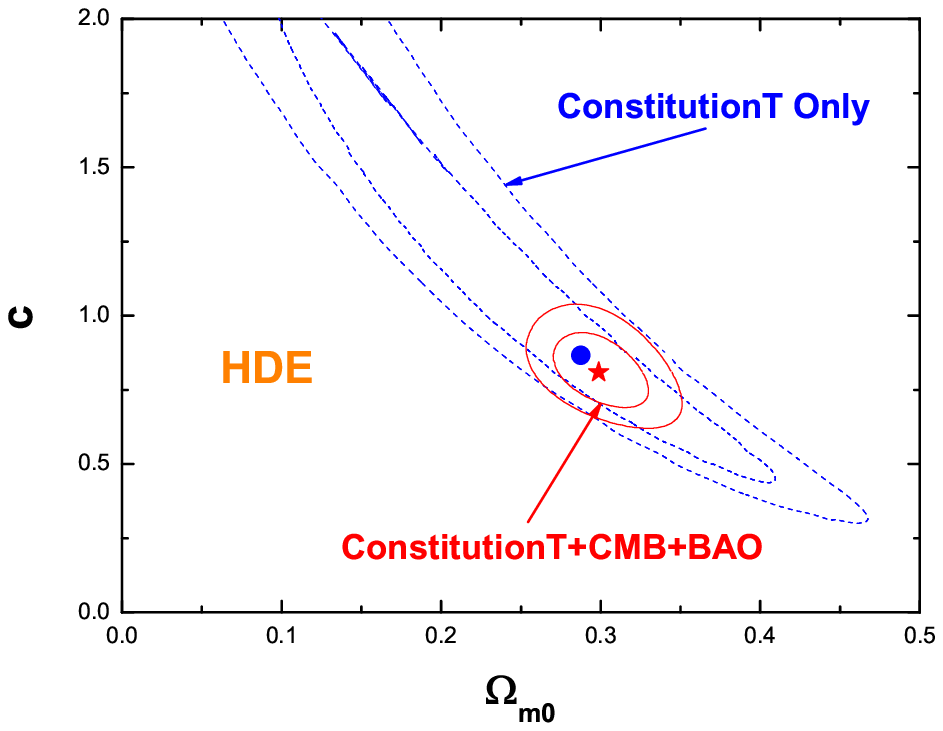}
\caption{\label{fig2}  The same as in Fig.\ref{fig1}, except for the cases of the Constitution and the ConstitutionT datasets.
It should be mentioned that,
in the left panel, the best-fit point for the SNIa only is outside the $2\sigma$ confidence region given by the combined SNIa+BAO+CMB data,
while in the right panel, the best-fit point for the SNIa only is very close to the best-fit point for the combined SNIa+BAO+CMB data.
This means that the ConstitutionT dataset is very helpful to remove the tension between SNIa data and other cosmological observations.}
\end{figure}

In \cite{wei10}, by performing the best-fit analysis on the CPL
model, Wei argued that the UnionT and the ConstitutionT datasets are
fully consistent with the other cosmological observations. To verify
this conclusion, more theoretical models should be taken into
account. Here we study the HDE model as an example. In
Fig.\ref{fig1}, we plot the $1\sigma$ and the $2\sigma$ confidence
level (CL) contours for the HDE model, where the Union and the
UnionT datasets are used, respectively. From this figure, we find
that the best-fit point for the Union data is outside the $2\sigma$
confidence region given by the combined Union+BAO+CMB data, while
the best-fit point for the UnionT data is inside the $1\sigma$
confidence region given by the combined UnionT+BAO+CMB data. This
means that the UnionT dataset is very useful to remove the tension
between SNIa data and other cosmological observations. In addition,
we also use the Constitution and the ConstitutionT datasets to plot
the CL contours for the HDE model in Fig.\ref{fig2}. It is found
that the best-fit point for the Constitution data is outside the
$2\sigma$ confidence region given by the combined
Constitution+BAO+CMB data, while the best-fit point for the
ConstitutionT data is very close to the best-fit point for the
combined ConstitutionT+BAO+CMB data. This means that the
ConstitutionT dataset is also very helpful to remove the tension.
Therefore, we conclude that the truncated SNIa datasets can remove
the tension between SNIa data and other cosmological observations.

(2) The CMB data is very helpful to break the degeneracy among different parameters,
and plays a very important role in distinguishing different DE models.

\begin{figure}
\includegraphics[scale=0.7, angle=0]{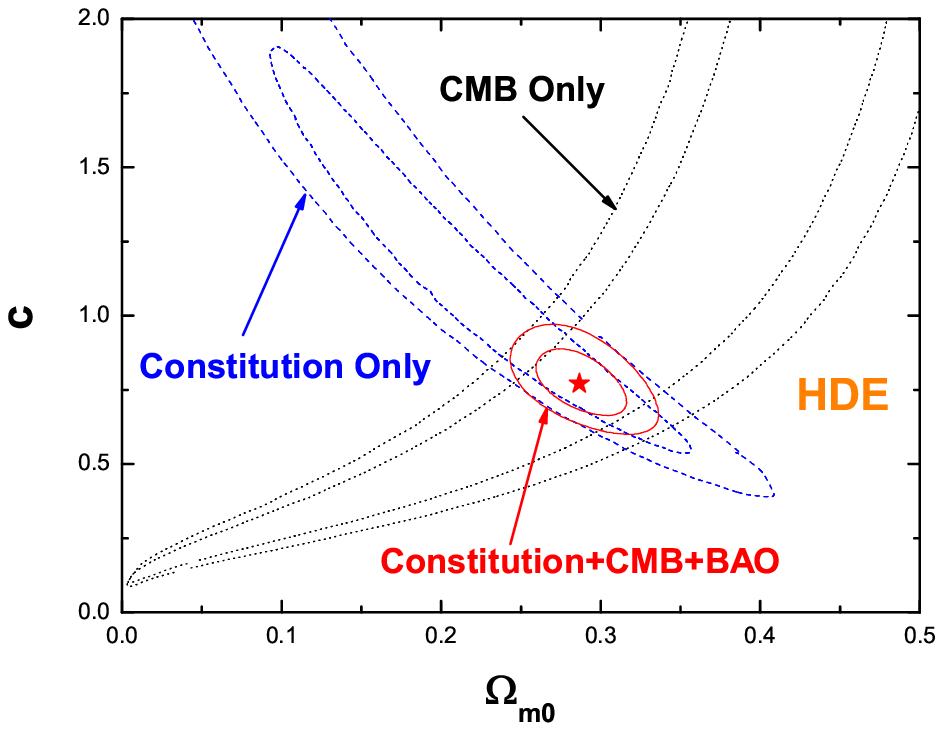}
\includegraphics[scale=0.7, angle=0]{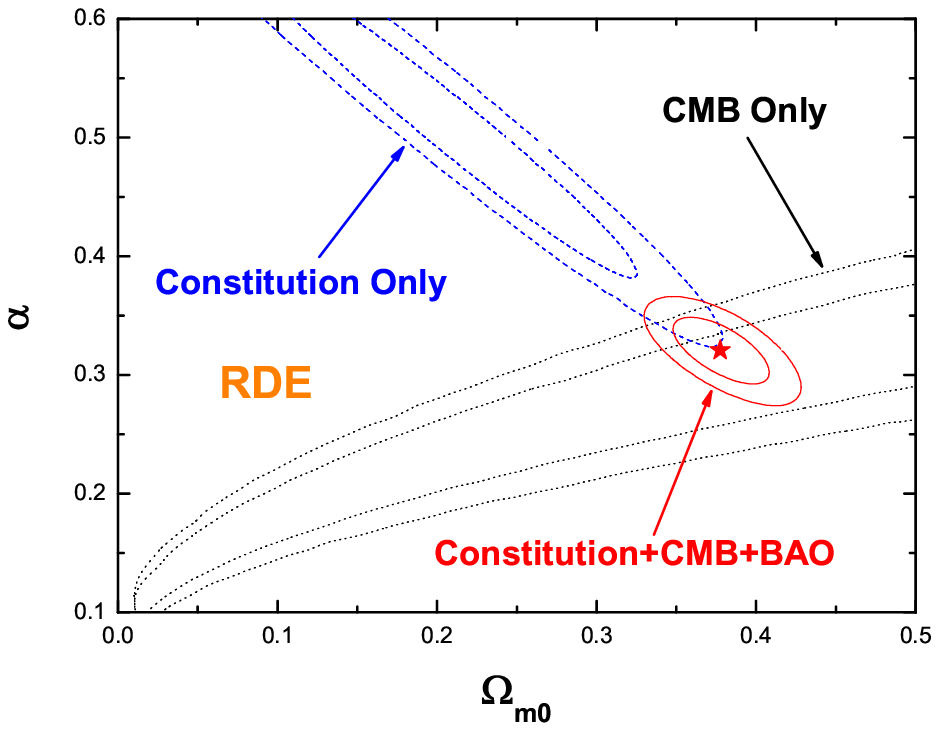}
\caption{\label{fig3} The $1\sigma$ and the $2\sigma$ CL contours for the HDE and the RDE model.
The Constitution dataset is used to plot this figure.
The left panel is plotted by using the HDE model, while the right panel is plotted by using the RDE model.
For both these two panels,
The blue dashed lines correspond to the constraints given by the SNIa data only,
the black dotted lines correspond to the constraints given by the CMB data only,
the red solid lines correspond to the constraints given by the combined SNIa+BAO+CMB data,
and the red stars denote the best-fit point for the combined SNIa+BAO+CMB data.
Since the shapes of CL contours given by the CMB data alone are quite different from that given by the SNIa data alone,
the CMB data is very helpful to break the degeneracy among different parameters.
As shown in the left panel, the $1\sigma$ CL contour of the HDE model given by the CMB data
intersects to the $1\sigma$ CL contour of the HDE model given by the SNIa data;
while in the right panel, the $1\sigma$ CL contour of the RDE model given by the CMB data
does not intersect to the $1\sigma$ CL contour of the RDE model given by the SNIa data.
This fact explains why the HDE model performs much better in fitting the combined Constitution+BAO+CMB data than the RDE model.
Besides, we also check the effect of the BAO data,
and find that the $1\sigma$ confidence region of the HDE model given by the BAO data alone can completely cover that given by the SNIa data.
For simplicity, the CL contours given by the BAO data are not plotted in this figure.}
\end{figure}

As an example, we will compare the HDE model with the RDE model.
By using the Constitution SNIa data alone, the CMB data alone, and the combined Constitution+CMB+BAO data, respectively,
we plot the $1\sigma$ and the $2\sigma$ CL contours for the HDE and the RDE model in Fig.\ref{fig3}.
From this figure,
we find that the shapes of CL contours given by the CMB data alone are quite different from that given by the SNIa data alone,
and the CMB data is very helpful to break the degeneracy among different parameters.
As shown in the left panel, the $1\sigma$ CL contour of the HDE model given by the CMB data
intersects to the $1\sigma$ CL contour of the HDE model given by the SNIa data;
while in the right panel, the $1\sigma$ CL contour of the RDE model given by the CMB data
does not intersect to the $1\sigma$ CL contour of the RDE model given by the SNIa data.
This fact explains why the HDE model performs much better in fitting the combined Constitution+BAO+CMB data than the RDE model.
Besides, we also check the effect of the BAO data,
and find that the $1\sigma$ confidence region of the HDE model given by the BAO data alone can completely cover that given by the SNIa data.
So the constraint given by the BAO data is quite weaker than that given by the SNIa and the BAO data.
This conclusion can be further verified by using table \ref{table3}.
As seen in table \ref{table3}, without adopting the CMB data, it is very difficult to distinguish the HDE Model from the RDE model.
After adding the CMB data,
it is seen that the $\chi_{min}^{2}$ of the HDE Model given by the SNIa+CMB data
is 6.811 smaller than the $\chi_{min}^{2}$ of the RDE Model given by the SNIa+CMB data,
while the $\chi_{min}^{2}$ of the HDE Model given by the combined SNIa+CMB+BAO data
is 7.082 smaller than the $\chi_{min}^{2}$ of the RDE Model given by the combined SNIa+CMB+BAO data.

\begin{figure}
\includegraphics[scale=0.7, angle=0]{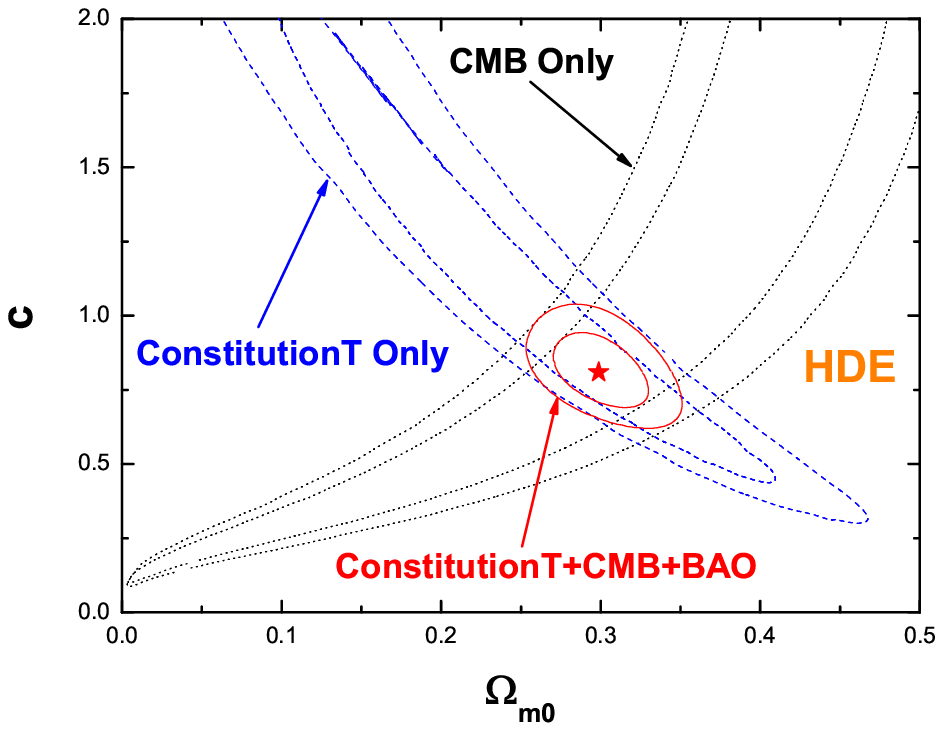}
\includegraphics[scale=0.7, angle=0]{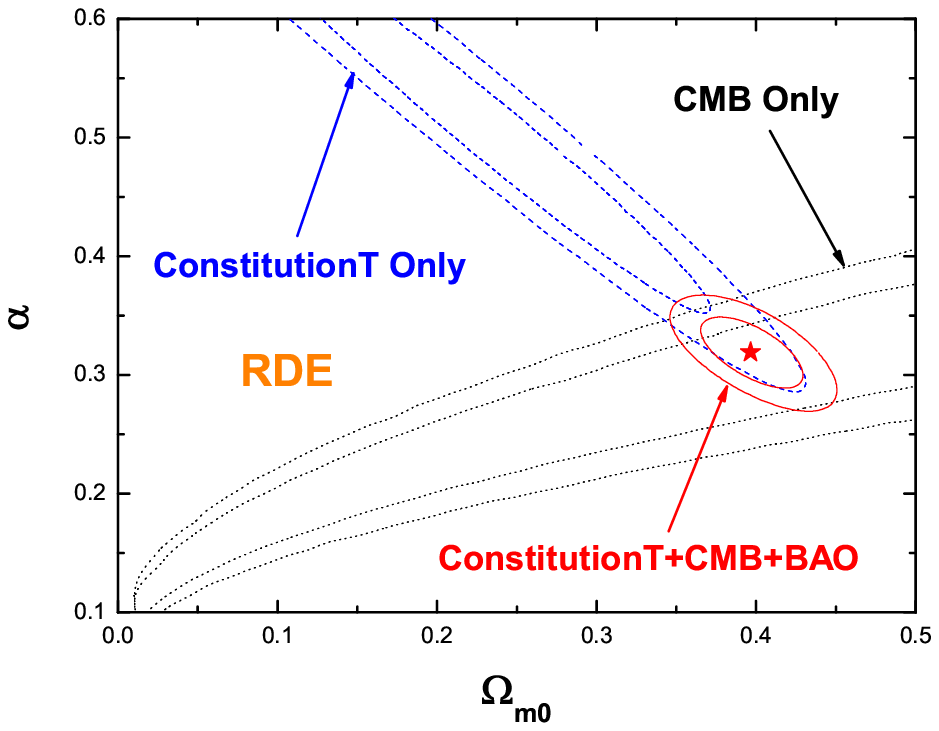}
\caption{\label{fig4} The same as in Fig.\ref{fig3}, except for the case of the ConstitutionT dataset.
Notice that the shapes of CL contours given by the CMB data alone are quite different from that given by the SNIa data alone.
As shown in the left panel, the $1\sigma$ CL contour of the HDE model given by the CMB data
intersects to the $1\sigma$ CL contour of the HDE model given by the SNIa data;
while in the right panel, the $1\sigma$ CL contour of the RDE model given by the CMB data
does not intersect to the $1\sigma$ CL contour of the RDE model given by the SNIa data.
This fact explains why the HDE model performs much better in fitting the combined ConstitutionT+BAO+CMB data than the RDE model.}
\end{figure}

In addition, by using the ConstitutionT SNIa data alone, the CMB data alone, and the combined ConstitutionT+CMB+BAO data, respectively,
we also plot the $1\sigma$ and the $2\sigma$ CL contours for the HDE and the RDE model in Fig.\ref{fig4}.
Again, we see that the $1\sigma$ CL contour of the HDE model given by the CMB data
intersects to the $1\sigma$ CL contour of the HDE model given by the SNIa data,
while the $1\sigma$ CL contour of the RDE model given by the CMB data
does not intersect to the $1\sigma$ CL contour of the RDE model given by the SNIa data.
This fact explains why the HDE model performs much better in fitting the combined ConstitutionT+BAO+CMB data than the RDE model.
Therefore, the CMB data is very helpful to break the degeneracy among different parameters,
and plays a very important role in distinguishing different DE models.

(3) The current observational data are still too limited to distinguish all DE models.

First, we discuss three kinds of two-parameter models: the XCDM model, the CG model, and the HDE model.
As seen in table \ref{table3},
the $\chi_{min}^{2}$ of these 3 models given by the combined Constitution+BAO+CMB data are 466.947, 466.891, and 467.384, respectively.
That is to say, after taking into account the CMB data, the differences of the $\chi_{min}^{2}$ of these 3 models are still smaller than 1.
Notice that this result also holds true in table \ref{table1}, table \ref{table2}, and  table \ref{table4}.
Therefore, one cannot judge which DE model is better.

Then, we discuss  three kinds of three-parameter models: the LP model, the CPL model, and the GCG model.
As seen in table \ref{table3},
the $\chi_{min}^{2}$ of these 3 models given by the combined Constitution+BAO+CMB data are 466.910, 466.902, 466.895, respectively,
and the differences of the $\chi_{min}^{2}$ of these 3 models are even smaller than 0.1.
Since this result also holds true in table \ref{table1}, table \ref{table2}, and  table \ref{table4},
it is also very difficult to distinguish these 3 models.
Therefore, to distinguish DE models better, more high-quality observational data are needed.

\section{Summary}

In this work, by performing the truncation procedure of \cite{wei10}
for 10 different models, we demonstrate that the approach adopted in
\cite{wei10} is valid. Moreover, by using the 4 SNIa datasets
mentioned above, as well as the observations of CMB and BAO, we
perform best-fit analysis on the 10 DE models. It is found that: (1)
For each DE model, the truncated SNIa datasets not only greatly
reduce $\chi _{min}^{2}$ and $\chi _{min}^{2}/dof$, but also remove
the tension between SNIa data and other cosmological observations.
(2) The CMB data is very helpful to break the degeneracy among
different parameters, and plays a very important role in
distinguishing different DE models. (3) The current observational
data are still too limited to distinguish all DE models. These
results provide a further support for Wei's work, and indicate that
the existence of biasing systematic errors in SNIa data should be
taken into account seriously.

\section*{Acknowledgements}
We are grateful to the reviewer for very useful and helpful suggestions.
We also thank Qing-Guo Huang, Hao Wei and Xin Zhang, for helpful discussions.
This work was supported by the NSFC grant
No.10535060/A050207, a NSFC group grant No.10821504 and Ministry of
Science and Technology 973 program under grant No.2007CB815401.
Shuang Wang was also supported by a graduate fund of USTC.


\end{document}